\documentclass[pra,aps,twocolumn,epsfig,superscriptaddress,showpacs]{revtex4}
\usepackage{amsmath}
\usepackage{amssymb}
\usepackage[dvips]{graphicx}%\usepackage{graphicx}% Include figure files
\usepackage{dcolumn}% Align table columns on decimal point
\usepackage{bm}% bold math
\usepackage[colorlinks=false,dvipdfm]{hyperref} % citecolor=blue 文件内部引用
\usepackage{setspace}
\usepackage{amsfonts}
\usepackage{rotating}
\usepackage{makeidx}
\usepackage{amsthm}

\newcommand{\tr}{\mbox{Tr}}

\begin{document}

\title{Bounds of concurrence and their relation with fidelity and frontier
states\footnote{Physics Letters A 373 (2009) 1616-1620}}

%\author{Zhihao Ma$^{1}$, Fu-Lin Zhang$^{2}$, Dong-Ling Deng$^{2}$, and Jing-Ling Chen$^{2}$$^{*}$}
%\address{$^{1}$Department of Mathematics, Shanghai Jiaotong
%University, Shanghai, 200240, P.R.China \\ $^{2}$Theoretical Physics
%Division, Chern Institute of Mathematics, Nankai University, Tianjin
%300071, People's Republic of China } \ead{{*} chenjl@nankai.edu.cn }

\author{Zhihao Ma}
%\email[Email:]{mazhihao@sjtu.edu.cn}
\affiliation{Department of Mathematics, Shanghai Jiaotong
University, Shanghai, 200240, P.R.China}

\author{Fu-Lin Zhang}
%\email[Email:]{flzhang@mail.nankai.edu.cn}
\affiliation{Theoretical
Physics Division, Chern Institute of Mathematics, Nankai University,
Tianjin, 300071, P.R.China
\\PHONE: 011+8622-2350-9287,
FAX: 011+8622-2350-1532}

\author{Dong-Ling Deng}
\affiliation{Theoretical Physics Division, Chern Institute of
Mathematics, Nankai University, Tianjin, 300071, P.R.China
\\PHONE: 011+8622-2350-9287,
FAX: 011+8622-2350-1532}

\author{Jing-Ling Chen}
\email[Email:]{chenjl@nankai.edu.cn}\affiliation{Theoretical Physics
Division, Chern Institute of Mathematics, Nankai University,
Tianjin, 300071, P.R.China
\\PHONE: 011+8622-2350-9287,
FAX: 011+8622-2350-1532}

%\date{\today}

\begin{abstract}
The bounds of concurrence in [F. Mintert and A. Buchleitner, Phys.
Rev. Lett. 98 (2007) 140505] and [C. Zhang \textit{et. al.}, Phys.
Rev. A 78 (2008) 042308] are proved by using two properties of the
fidelity. In two-qubit systems, for a given value of concurrence,
the states achieving the maximal upper bound, the minimal lower
bound or the maximal difference upper-lower bound are determined
analytically.
%The bounds of concurrence in [F. Mintert and A. Buchleitner, Phys.
%Rev. Lett. 98 (2007) 140505] and [C. Zhang \textit{et. al.}, arXiv:
%0806.2598] are proved by using two properties of the fidelity. This
%shows these two concepts in quantum information are closely related.
\end{abstract}

\pacs{03.67.-a, 03.65.Ta}

% \keywords{Concurrence; Fidelity;
%Frontier states}

 \maketitle

\section{Introduction\label{Intro}}
 Entanglement, which depicts the
nonclassical connection between two parts of a quantum system, has
been recognized as crucial in various field of quantum information
in recent years \cite{Book,teleport,code,comput,comput1,key,key1}.
Several measures have been proposed to quantify the degree to which
a state is entangled, such as entanglement of formation
\cite{EOF,Wootters97,Wootters98}, entanglement of distillation
\cite{EOD}, relative entropy of entanglement \cite{REE}, negativity
\cite{NEG,NEG1}, and so on. For two-qubit systems, the entanglement
of formation is equivalent to a computable quantity, which is
referred to as \textit{concurrence} \cite{Wootters97,Wootters98}.
% Entanglement and fidelity are two important concepts in
%quantum information theory \cite{Book}.  The former depicts the
%nonclassical connection between two parts of a quantum system. And
%the latter is a measure of \textit{closeness} of two quantum states.
%They are shown to be closely related to each other in many aspects
%\cite{Uhlmann2000,Wootters97,Guo,Wootters98}.
%One of the most famous measures of entanglement is the concurrence
%\cite{Wootters97,Wootters98} of two-qubit system, which is
%equivalent to the entanglement of formation \cite{EOF}.
The concurrence of a pure two-qubit state $| \psi \rangle$ is given
by
\begin{eqnarray} \label{ConP}
C(| \psi \rangle) =\sqrt{2(1- \tr {\rho^A}^2)}= \sqrt{2(1- \tr
{\rho^B}^2)},
\end{eqnarray}
where $\rho^A= \tr_B | \psi \rangle \langle \psi |$ is the partial
trace of $| \psi \rangle \langle \psi |$ over subsystem $B$, and
$\rho^B$ has a similar meaning. For a mixed state, the concurrence
is defined as the average concurrence of the pure states of the
decomposition, minimized over all decompositions of $\rho =
\sum_{j}p_{j} | \psi_j \rangle  \langle \psi_j |$,
\begin{eqnarray}\label{ConM}
C(\rho)= \min \sum_{j}p_{j}C(| \psi_j \rangle).
\end{eqnarray}
%The decomposition corresponding to the minimum in Eq. (\ref{ConM})
%is called the optimal decomposition.
It has been proved in \cite{Wootters97,Wootters98} that Eq.
(\ref{ConM}) can be expressed explicitly as
\begin{eqnarray}\label{ConEx}
C(\rho)= \max
\{0,\sqrt{\lambda_{1}}-\sqrt{\lambda_{2}}-\sqrt{\lambda_{3}}-\sqrt{\lambda_{4}}\},
\end{eqnarray}
in which $\lambda_{1},...,\lambda_{4}$ are the eigenvalues of the
operator $R=\rho (\sigma_{y} \otimes \sigma_{y} ) \rho^{*}
(\sigma_{y} \otimes \sigma_{y} )$ in decreasing order and
$\sigma_{y}$ is the second Pauli matrix.
%The decomposition making the convex combination reach the minimum is
%called an optimal one.

Recently, many works about detecting entanglement experimentally
have been reported
\cite{Terhal00,Mintert,Mintert05,Eisert07,ZhangCJ,NiuXL}. In one of
the existing schemes, the authors of \cite{Mintert,ZhangCJ}
presented observable lower and upper bounds of the squared
concurrence
\begin{eqnarray} \label{Bounds}
2 \bigr[ \tr\rho^{2}-\tr{\rho^A}^{2} \bigr] \leq C^{2}(\rho) \leq 2
\bigr[1-\tr{\rho^A}^{2} \bigr],
\end{eqnarray}
where $C(\rho)$ is the concurrence for \textit{arbitrary
dimensional} states, taking the definitions in Eq. (\ref{ConP}) and
(\ref{ConM}). These bounds can be easily obtained by a few
experimental measurements on a \textit{twofold copy} $\rho \otimes
\rho$ of the mixed states.

It is shown that the bounds provide an excellent estimation of
concurrence for weakly mixed (called \textit{quasipure} in
\cite{Mintert}) states  \cite{ZhangCJ}, and they have certain
relations with the degree of mixing for a mixed state and some
properties of the linear entropy. In this letter, we give closer
analysis of the bounds. First, we find the inequality (\ref{Bounds})
can be distinctly understood in the viewpoint of fidelity, which is
another important concept in quantum information
\cite{Book,Bures,Uhlmann,Uhlmann1}. The details are given in Sec.
\ref{proof}. To show the departure of the bounds for a given
concurrence, in Sec. \ref{Frontier}, the two-qubit frontier states
are determined analytically. Namely, we present the quantum states,
which achieve the maximal upper bound, the minimal lower bound or
the maximal difference upper-lower bound, for a given value of
entanglement. In Sec. \ref{Concl}, we discuss the relations of the
bounds and concurrence or the linear entropy, and give a brief
conclusion.

\section{Proof by fidelity\label{proof}}
Fidelity is a measure of \textit{closeness} of two quantum states,
which is shown to be closely related to entanglement in many aspects
\cite{Uhlmann2000,Wootters97,Guo,Wootters98}. The fidelity between
two states $\rho_1$ and $\rho_2$ of a quantum system $R$ reads
\cite{Book,Bures,Uhlmann,Uhlmann1}
\begin{eqnarray}
F(\rho_1,\rho_2)= \biggr[\tr \biggr( \sqrt{\sqrt{\rho_1} \rho_2
\sqrt{\rho_1}} \biggr) \biggr]^2.
\end{eqnarray}
When $\rho_1=| \phi \rangle\langle \phi |$ and $\rho_2= | \varphi
\rangle\langle \varphi |$ are two pure states, $F(\rho_1,\rho_2)=\tr
(| \phi \rangle  \langle \phi | \varphi \rangle  \langle \varphi |)
= | \langle \phi | \varphi \rangle |^2$. Let $Q$ be a copy of $R$,
and $| \psi_1 \rangle$ is a purification of $ \rho_1 $ and $| \psi_2
\rangle$ of $\rho_2$ into $RQ$, then
\begin{eqnarray}
F(\rho_1,\rho_2) \geq  F(\sigma_1,\sigma_2)= | \langle \psi_1 |
\psi_2 \rangle |^2,
\end{eqnarray}
where $\sigma_1=| \psi_1 \rangle  \langle \psi_1 |$, $\sigma_2=|
\psi_2 \rangle \langle \psi_2 |$, and $\tr_Q \sigma_1 = \rho_1$,
$\tr_Q \sigma_2 = \rho_2$. It is proved in \cite{SuperF}, that
fidelity can be bounded above by the so called super-fidelity
$G(\rho_1, \rho_2)$:
\begin{eqnarray}
F(\rho_1, \rho_2) &\leq& G(\rho_1, \rho_2) \nonumber \\
&=&\tr \rho_1 \rho_2 + \sqrt{(1-\tr \rho_1^2)(1-\tr \rho_2^2)} \nonumber\\
&\leq& 1 .
\end{eqnarray}
When $\rho_1$ and $\rho_2$ are two pure states, $G(\rho_1,
\rho_2)=\tr \rho_1 \rho_2 =F(\rho_1, \rho_2)$. The following is the
proof of the inequality (\ref{Bounds}) by using the properties of
fidelity and super-fidelity mentioned above.

\emph{Proof.} Suppose $\{ | \psi_i \rangle \}$ is an decomposition
 minimizing the average concurrence in Eq. (\ref{ConM}),
$\rho_i= | \psi_i \rangle \langle \psi_i | $, and $\rho= \sum_i t_i
\rho_i$, where $\sum_i t_i =1$. The reduced density operator
$\rho^A= \sum_i t_i \rho_i^A$, where $\rho_i^A= \tr_B \rho_i$. It is
straightforward to obtain
\begin{eqnarray}
C^2(\rho)&=& 2 \sum_{i,j} t_i t_j \sqrt{ \bigr[1- \tr {\rho_i^A}^2
\bigr] \bigr[1- \tr {\rho_j^A}^2 \bigr]}, \nonumber \\
\tr \rho^2 &=& \sum_{i,j} t_i t_j \tr(\rho_i \rho_j), \\
\tr {\rho^A}^2 &=& \sum_{i,j} t_i t_j \tr(\rho_i^A \rho_j^A).
\nonumber
\end{eqnarray}
Then, one can notice that
\begin{eqnarray}
1 &\geq& \tr(\rho_i^A \rho_j^A)+\sqrt{ \bigr[1- \tr {\rho_i^A}^2
\bigr] \bigr[1- \tr {\rho_j^A}^2 \bigr]} \nonumber \\ &=&
G(\rho_i^A,\rho_j^A)
 \geq
F(\rho_i^A,\rho_j^A).
\end{eqnarray}
On the other hand, $\rho_i$ and $\rho_j$ are two purifications of
$\rho_i^A$ and $\rho_j^A$, we obtain
\begin{eqnarray}
1  \geq G(\rho_i^A,\rho_j^A)
 \geq F(\rho_i^A,\rho_j^A)  \geq F(\rho_i,\rho_j)= \tr (\rho_i \rho_j).
\end{eqnarray}
Hence,
\begin{eqnarray}
2 \geq C^2(\rho)+2\tr {\rho^A}^2 &=& 2 \sum_{i,j} t_i t_j
G(\rho_i^A,\rho_j^A)\nonumber \\
&\geq& 2 \sum_{i,j} t_i t_j \tr (\rho_i \rho_j)= 2\tr {\rho}^2. \ \
\
\end{eqnarray}
This ends the proof.

%The above result shows that, the properties of the fidelity between
%two states in the optimal decomposition of a biparticle system lead
%to the upper and lower bounds of concurrence.
%The bounds are closely
%related to the schemes to observe entanglement experimentally. We
%can foretell there is more profound connection between entanglement
%and fidelity to be discovered.

\section{Frontier states\label{Frontier}}
In the above section, we show the properties of the fidelity lead to
the upper and lower bounds of squared concurrence in Eq.
(\ref{Bounds}). For the sake of brevity, we adopt the
\textit{tangle} \cite{Tangle} to replace concurrence, $\tau = C^2$.
%Considering the different between the subsystems in a biparticle
%system, the bounds of  $\tau$ should be written as
It is bounded by $\tau_{L} \leq  \tau  \leq \tau_{U}$, where
\begin{eqnarray} \label{BoundsT}
%&&\tau_{L}=\max \{2 \bigr[ \tr\rho^{2}-\tr{\rho^A}^{2} \bigr] ,2 \bigr[ \tr\rho^{2}-\tr{\rho^B}^{2} \bigr] \} , \nonumber \\
\tau_{L}&=&2  \tr\rho^{2}- 2 \min \{ \tr{\rho^A}^{2}, \tr{\rho^B}^{2}\}  , \nonumber \\
\tau_{U}&=&2- 2 \max \{ \tr{\rho^A}^{2}, \tr{\rho^B}^{2}\}.
\end{eqnarray}
The aim of this section is to derive the quantum states achieving
the maximal $\tau_{U}$, the minimal $\tau_{L}$ or the maximal their
difference $\Delta = \tau_{U}-\tau_{L}$, for a given value of
$\tau$. These frontier state are denoted as $\rho_{U}$, $\rho_{L}$
and $\rho_{\Delta}$ respectively in the following paragraphes. We
only consider the two-qubit case, for which the concurrence has the
exact formula as Eq. (\ref{ConEx}).

\subsection{Frontier state of $\tau_{U}$, $\rho_{U}$}
It is easy to obtain, when the purities of the two subsystem
$\tr{\rho^A}^{2} = \tr{\rho^B}^{2} =\frac{1}{2}$, $\tau_{U}=1$
achieves the maximum. A straightforward choice of the frontier of
$\tau_{U}$ is the isotropic state or say the Werner state
\cite{IsoState,WState}
\begin{eqnarray} \label{RhoW}
\rho_{U}=\rho_{W}=\frac{1-x}{4}I_{2} \otimes I_{2} + x
|\Phi_{+}\rangle \langle\Phi_{+}|,
\end{eqnarray}
where $I_{2}$ is the $2 \times 2$ identity matrix and
$|\Phi_{+}\rangle = [|00\rangle+|11\rangle]/\sqrt{2}$ is one of the
Bell basis. Then $\tau(\rho_{U})={\bigr[\max
\{0,\frac{3}{2}(x-\frac{1}{3}) \}\bigr]}^2$ and
$\tau_{U}(\rho_{U})=1$ for any value of tangle.

\subsection{Frontier state of $\Delta$,  $\rho_{\Delta}$}
The difference between the upper and lower bounds can be written as
\begin{eqnarray} \label{delt}
\Delta = 2 \bigr[1-\tr{\rho}^{2} \bigr] -
2\bigr|\tr{\rho^A}^{2}-\tr{\rho^B}^{2}\bigr|.
\end{eqnarray}
The first part of Eq. (\ref{delt}) is the linear entropy \cite{LE},
a measure for the degree of of mixture of the quantum state. In
\cite{MEMS}, the authors present a class of \textit{maximally
entangled mixed states}:
\begin{eqnarray} \label{RhoMems}
\rho_{MEMS}=\begin{bmatrix}
 \ g(\gamma) & 0 & 0 & \gamma/2\\
 \ 0 & 1-2g(\gamma) & 0 & 0\\
 \ 0 & 0 & 0 & 0\\
 \ \gamma/2 & 0 & 0 & g(\gamma)
\end{bmatrix},
\end{eqnarray}
with $g(\gamma)=1/3$ for $0 \leq \gamma \leq 2/3$ and
$g(\gamma)=\gamma/2$ for $2/3 \leq \gamma \leq 1$. For a given value
of tangle, its linear entropy
\begin{eqnarray} \label{LEMax}
2 \bigr[1-\tr{\rho_{MEMS}}^{2} \bigr]=4
g(\gamma)\bigr[2-3g(\gamma)\bigr]-\gamma^2,
\end{eqnarray}
reaches the maximum over all the two-qubit states, with
$\tau=\gamma^2$. Straightforward calculations indicate the two
subsystems have the equivalent purities $\tr{\rho_{MEMS}^A}^{2} =
\tr{\rho_{MEMS}^B}^{2} =2 g^2(\gamma) -2 g(\gamma) +1$. Therefore,
the state in Eq. (\ref{RhoMems}) makes $\Delta$ maximal,
$\rho_{\Delta}=\rho_{MEMS}$, and the corresponding maximum equals
the linear entropy in Eq. (\ref{LEMax}).

\subsection{Frontier state of $\tau_{L}$, $\rho_{L}$}

\begin{figure}
\includegraphics[width=7cm]{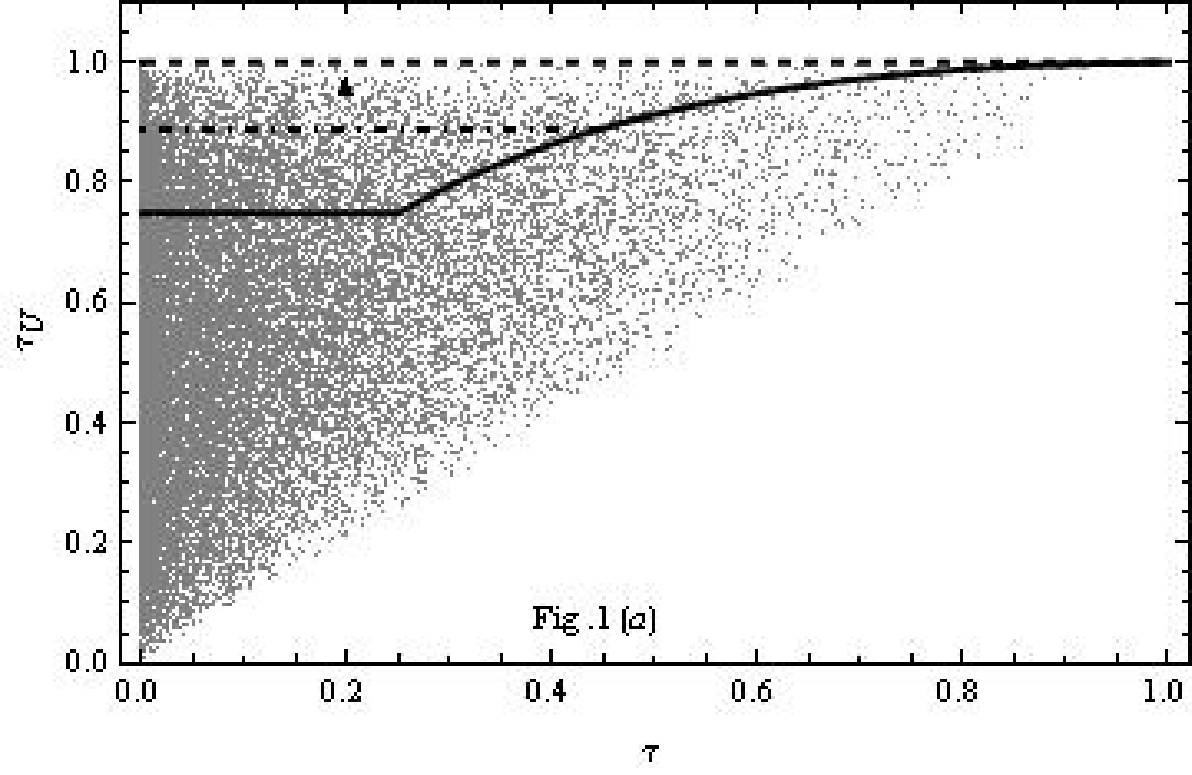}
\includegraphics[width=7cm]{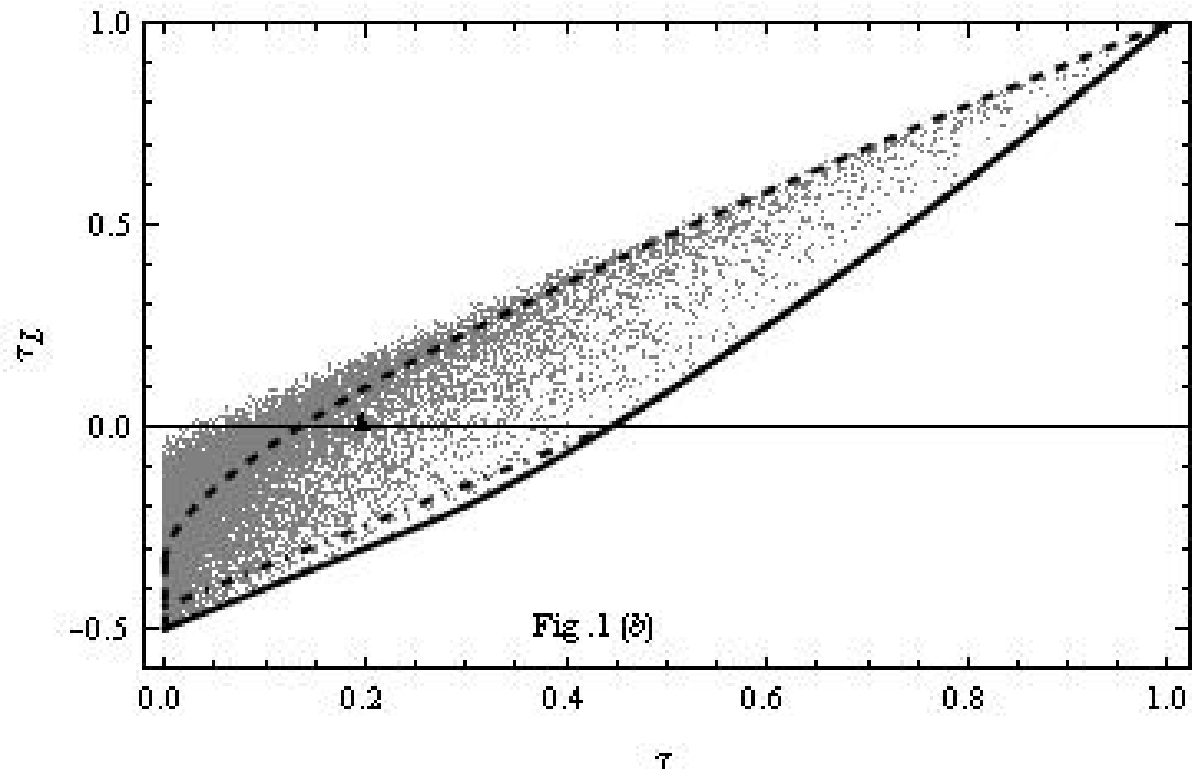}
\includegraphics[width=7cm]{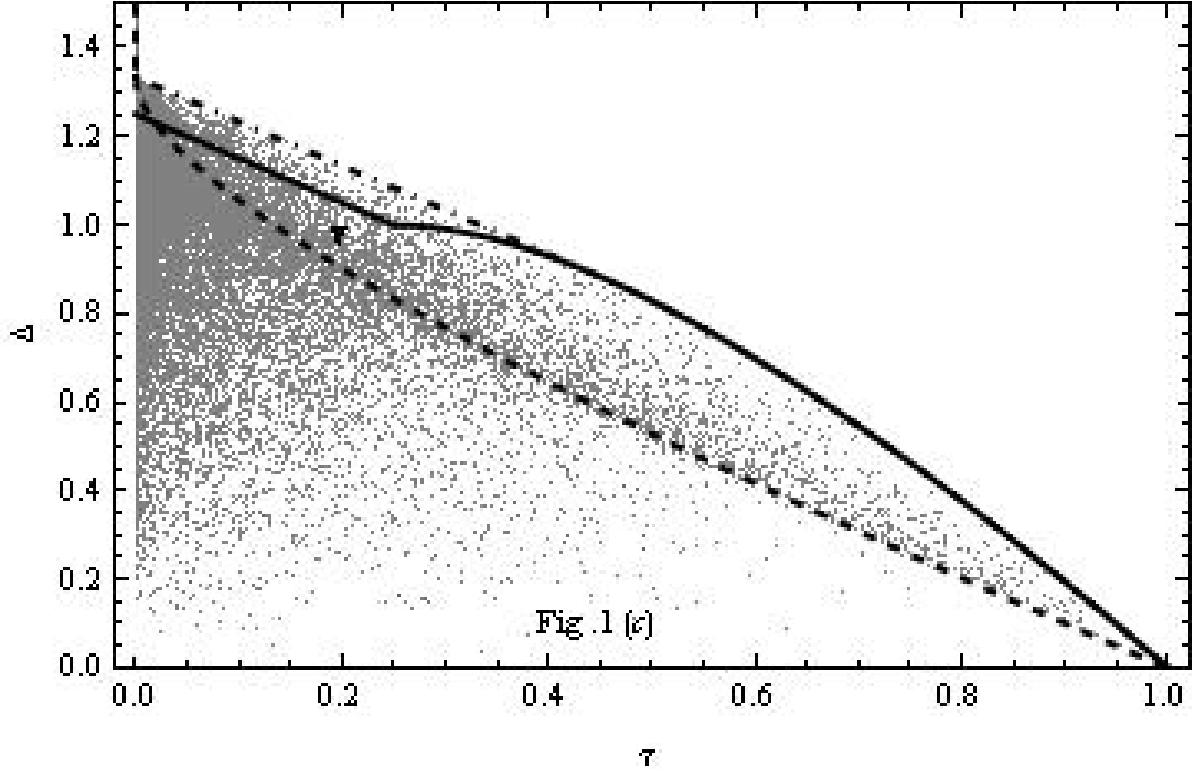}
 \caption{Plot of (a) $30000$ randomly generated states in the $\tau_{U}$-$\tau$ plane;
 (b) and (c) $30000$ randomly generated states weighted in the $\tau_{L}$-$\tau$ and $\Delta$-$\tau$ planes.
 The curves in the planes are $\rho_{U}$ (dashed), $\rho_{L}$ (solid) and $\rho_{\Delta}$ (dot-dashed).
 The small triangles in the planes indicate an intermediate state between $\rho_{U}$ and $\rho_{\Delta}$.}
\label{fig1}
\end{figure}

The region of the two-qubit states in the $\tau_{L}$-$\tau$ plane is
shown in Fig. \ref{fig1}(b). The state $\rho_{L}$, with the minimal
$\tau_{L}$ for a given $\tau$, achieves the maximal $\tau$ when the
value of $\tau_{L}$ is fixed. To derive $\rho_{L}$, we postulate it
take the ansatz form
\begin{eqnarray} \label{RhoAnsatz}
\rho_{ansatz}=\begin{bmatrix}
 \ x+\gamma/2 & 0 & 0 & \gamma/2\\
 \ 0 & a & 0 & 0\\
 \ 0 & 0 & b & 0\\
 \ \gamma/2 & 0 & 0 & y+\gamma/2
\end{bmatrix},
\end{eqnarray}
with the non-negative real parameters, $x+y+a+b+\gamma=1$. It
comprises a mixture of the Bell state $| \Phi_{+} \rangle$ and a
mixed diagonal state. There are two reasons for choosing this ansatz
state. (\romannumeral1) The frontier states $\rho_{U}$ and
$\rho_{\Delta}$ determined above are relating to this form of
$\rho_{ansatz}$. As shown in Fig. \ref{fig1}(a), when the tangle
takes a large value, for instance $0.8$ or larger, $\tau_{U}$ of
$\rho_{\Delta}$ is close to $1$, moreover its $\Delta$ is the
maximum. This indicates $\rho_{\Delta}$ is a candidate for the
frontier $\rho_{L}$. (\romannumeral2) For a pure state, the lower
bound $\tau_{L}=\tau$. When the state is mixed, $\tau_{L}$ deviates
from the tangle $\tau$. This can be shown by a special type states
in Fig. \ref{fig3}(d), which will be discussed in Sec. \ref{Concl}.
Therefore, we can speculate the value of $\tau_{L}$  is a token of
purity of a state. Otherwise, all the maximally entangled states for
a given degree of mixing derived in \cite{MEMS,MEMS1} take the form
in Eq. (\ref{RhoAnsatz}), with various combinations of entanglement
and mixedness measures.

The tangle of $\rho_{ansatz}$ is given by
\begin{eqnarray}\label{tangleA}
\tau= \bigr\{ \max \{ \gamma- 2\sqrt{ab}, 0 \} \bigr\}^2,
\end{eqnarray}
with the maximum $\tau=\gamma^2$ when $b=0$, as shown in
\cite{MEMS}. One can surmise directly $\tr{\rho^A}^{2} =
\tr{\rho^B}^{2}$, when $\tau_{L}$ reached its minimum for a fixed
$\tau$, which requires $x=y=(1-a-\gamma)/2$. Then, the lower bound
is given by
\begin{eqnarray}\label{TauL}
\tau_{L}= 2 a^2 -2a + \gamma^2,
\end{eqnarray}
with $a \in [0,1-\gamma]$. The upper bound of $a$ holds $x=y \geq
0$. Therefore, when $1-\gamma \geq 1/2$, or say $\gamma \leq 1/2$,
the minimum points of $\tau_{L}$ are $a=1/2$ and $x=(1-2\gamma)/4$.
In the other region  $\gamma \geq 1/2$, the optimal solution occurs
when $a=1-\gamma$ and $x=0$. Thus, the frontier state, achieving the
minimal $\tau_{L}$ for a fixed $\tau$, can be written in the same
form as  $\rho_{\Delta}$
\begin{eqnarray} \label{RhoL}
\rho_{L}=\begin{bmatrix}
 \ f(\gamma) & 0 & 0 & \gamma/2\\
 \ 0 & 1-2f(\gamma) & 0 & 0\\
 \ 0 & 0 & 0 & 0\\
 \ \gamma/2 & 0 & 0 & f(\gamma)
\end{bmatrix},
\end{eqnarray}
with $ f(\gamma)=1/4$ for $0 \leq \gamma \leq 1/2$ and $
f(\gamma)=\gamma/2$ for $1/2 \leq \gamma \leq 1$. For a given degree
of entanglement $\tau=\gamma^2$, its $\tau_{L}$ is given by
\begin{eqnarray}
\tau_{L}= \left \{
\begin{array}{lr}
\gamma^2-1/2,      \; &  0 \leq \gamma \leq 1/2 \;;\\
3\gamma^2-2\gamma, \; &  1/2 \leq \gamma \leq 1 \;.
%\gamma^2-\frac{1}{2}, \; & 0 \leq \gamma \leq \frac{1}{2}\;\\
%3\gamma^2-2\gamma, \;& \frac{1}{2} \leq \gamma \leq 1\;,
\end{array}
\right.
\end{eqnarray}
To verify the frontier state $\rho_{L}$, we randomly generate one
million two-qubit matrices, which indicate the region of physically
acceptable states in the $\tau_{L}-\tau$ plane is encircled
perfectly by the curve of the state in Eq. (\ref{RhoL}). We plot the
curve comparing with $30000$ weighted random states, namely the
mixtures of random states and $\rho_{L}$ with random weights, in
Fig. \ref{fig1}(b). The curve joins with the one of $\rho_{U}$ at
$(\tau, \tau_{L})=(0,-1/2)$, where $\rho_{L}=\bigr[ |00\rangle
\langle 00| + |11\rangle \langle 11| + 2|01\rangle \langle 01|
\bigr]/4$ and $\rho_{U}= I_{2} \otimes I_{2}/4$. When $\gamma \geq
2/3$, $\rho_{L}=\rho_{\Delta}$; and when $\gamma < 2/3$, the values
of $\tau_{L}$ for $\rho_{L}$ and $\rho_{\Delta}$ are less than $0$.

\subsection{Frontier state of $\Delta_{R}$, $\rho_{\Delta_{R}}$}

In the above discussion, we notice, when the degree of entanglement
$\tau \leq 4/9$, the lower bound $\tau_{L}$ has the possibility of
being less than $0$. But the value of concurrence, as a good defined
entanglement measure, should be non-negative
\cite{Vedral,Horodecki00}. This suggests a more reasonable lower
bound of the squared concurrence should be corrected as
\begin{eqnarray}\label{TauLR}
\tau_{L_{R}}=\max \{ \tau_{L},0  \},
\end{eqnarray}
whose difference with the upper bound is given by
\begin{eqnarray}\label{deltR}
\Delta_{R}=\tau_{U}-\tau_{L_{R}} =\left \{
\begin{array}{lr}
\tau_{U},      \; &  \tau_{L} < 0 \;;\\
\tau_{U}-\tau_{L}, \; &  \tau_{L} \geq 0 \;.
\end{array}
\right.
\end{eqnarray}
Then, both the frontier states $\rho_{L}$ and $\rho_{\Delta}$
correspond with the minimal $\tau_{L_{R}}$ for a given value of
$\tau$. A subsequent question arises: what is the frontier state
$\rho_{\Delta_{R}}$, which achieve the maximal $\Delta_{R}$ for a
given $\tau$?

There are two regions of $\tau$, where the frontier state
$\rho_{\Delta_{R}}$ can be determined directly. The first one occurs
for $\tau \leq \bigr( 2-\sqrt{3} \bigr) /2$, and the second for
$\tau \geq 4/9 $. In the former case, the $\tau_{L}$ of $\rho_{U}$
or say the Werner state is less than $0$, whose $\tau_{U}=1$. Hence,
 $\rho_{\Delta_{R}} =\rho_{U}= \rho_{W}$ in this
region. In the later case, $\rho_{\Delta_{R}} =\rho_{\Delta}=
\rho_{MEMS}$, since $\tau_{L} \geq 0$ for all the two-qubit states
with $\tau \geq 4/9 $.

In the in-between region with $ \bigr( 2-\sqrt{3} \bigr) /2 < \tau <
4/9$, $\tau_{L} (\rho_{W}) = (-1+2 \sqrt{\tau }+2 \tau )/3 > 0$ and
$\tau_{L} (\rho_{MEMS}) = -4/9+\tau  < 0$. Their upper bounds of
tangle are constants, $\tau_{U} (\rho_{W}) = 1$ and $\tau_{U}
(\rho_{MEMS}) = 8/9$. One can image an intermediate state, as shown
by the black triangles ($\blacktriangle$ and $\blacktriangledown$)
in Fig. \ref{fig1}, whose $\tau_{U}$ and $\tau_{L}$ are intermediate
between the ones of $\rho_{W}$ and $\rho_{MEMS}$. When it transforms
from $\rho_{MEMS}$ to $\rho_{W}$, as expressed by the directions of
the triangles, $\tau_{U}$ and $\tau_{L}$ are enhanced, whereas
$\Delta$ decreases. Considering the relations in Eq. (\ref{deltR}),
we can postulate the maximum of $\Delta_{R}$ occurs when
$\tau_{L}=0$. The Werner state and $\rho_{MEMS}$, in this region,
can be uniformed by
\begin{eqnarray} \label{RhoTriangle}
\nonumber \rho_{\blacktriangle}= \frac{1}{2}
\begin{bmatrix}
 \ 1-S & 0 & 0 & \gamma+\sqrt{S^2-M^2}\\
 \ 0 & S+M & 0 & 0\\
 \ 0 & 0 & S-M & 0\\
 \ \gamma+\sqrt{S^2-M^2} & 0 & 0 & 1-S
\end{bmatrix}, \ \
\end{eqnarray}
with the parameters $0 \leq M \leq S \leq 1$, $ \bigr( \sqrt{3}-1
\bigr) /2 < \gamma < 2/3$, and the tangle $\tau=\gamma^2$. When
$S=M=1/3$, $\rho_{\blacktriangle}=\rho_{MEMS}$; and when
$S=(1-\gamma)/3$ and $M=0$, $\rho_{\blacktriangle}$ returns into the
Werner state in term of the concurrence $\gamma$ \cite{MEMS1}.

Taking the intermediate state $\rho_{\blacktriangle}$ as a candidate
for $\rho_{\Delta_{R}}$, we will determine the values of $S$ and $M$
for a given $\gamma$ in the following. The bounds of tangle for the
state $\rho_{\blacktriangle}$ are given by
\begin{eqnarray}\label{BoundsTriangle}
\tau_{L}&=&-2S(1-S)+(\gamma+\sqrt{S^2-M^2})^2, \nonumber \\
\tau_{U}&=&1-M^2.
\end{eqnarray}
The condition $\tau_{L}=0$ leads to
\begin{eqnarray}\label{MS}
M= \mathcal{M}(S)= \sqrt{ S^2- \bigr[ \sqrt{2S(1-S)} -\gamma^2
\bigr]^2},
\end{eqnarray}
with $S \in [\mathcal{S}_{-}(\gamma),\mathcal{S}_{+}(\gamma)]$.
Here, $\mathcal{S}_{\pm}(\gamma)=(1 \pm \sqrt{1-2\gamma^2})/2$ keep
the value of $\sqrt{S^2-M^2}$ to be non-negative. Under such a
condition, the maximum of $\Delta_{R}$ occurs at
\begin{eqnarray}\label{SGamma}
S=\mathcal{S}(\gamma)= \left \{
\begin{array}{lr}
\mathcal{S}_{0}(\gamma),      \; &  \bigr( \sqrt{3}-1 \bigr) /2 < \gamma < \gamma_{c} \; \\
\mathcal{S}_{-}(\gamma), \; & \gamma_{c} < \gamma  < 2/3 \;,
\end{array}
\right.
\end{eqnarray}
where $\gamma_{c} \approx 0.38218$, and $\mathcal{S}_{0}(\gamma)$ is
the root of the equation $\partial_{S} \mathcal{M}(S)=0$ in the
region $[\mathcal{S}_{-}(\gamma),\mathcal{S}_{+}(\gamma)]$.

\begin{figure}
\includegraphics[width=7cm]{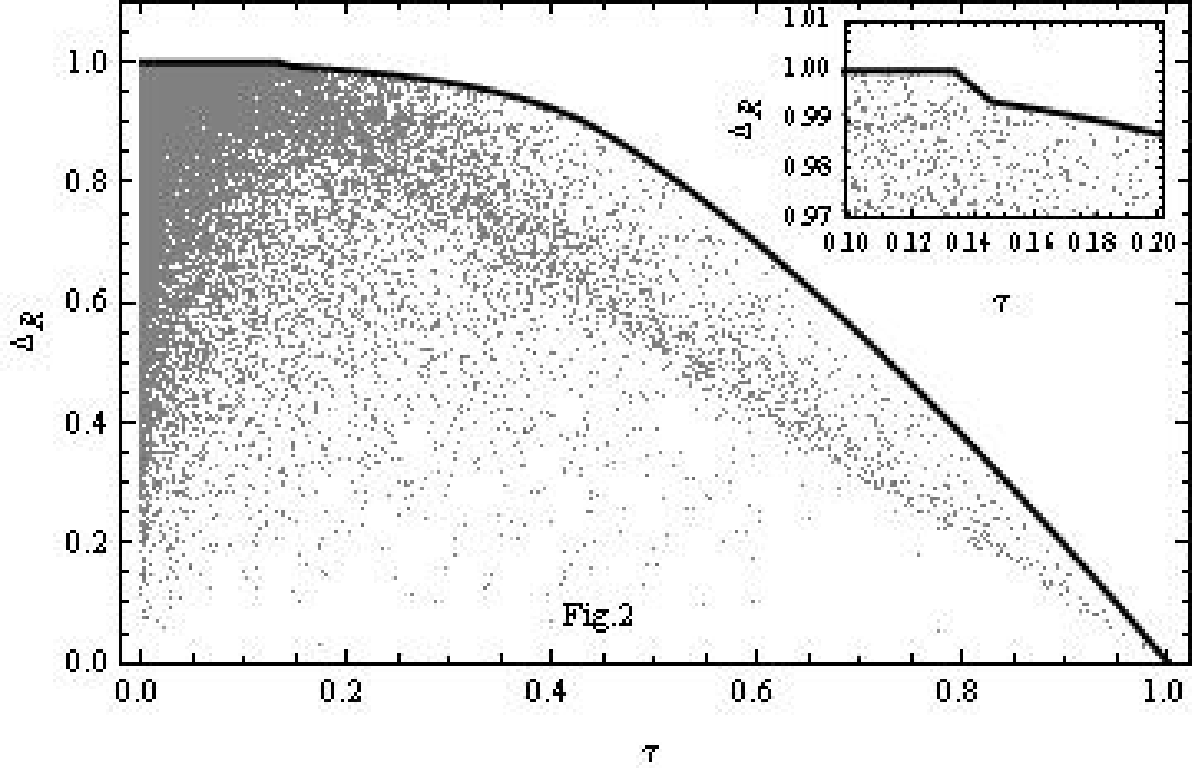} \\
 \caption{Curve of $\rho_{\Delta_{R}}$ is plotted in the $\Delta_{R}-\tau$ plane, in
company with 30000 random states weighted based on it. Upper right
is the magnified region nearby the two critical points, $\tau=\bigr(
2-\sqrt{3} \bigr) /2$ and $\tau=\gamma_{c}^2$} \label{fig2}
\end{figure}

Comparing the solutions of $\rho_{\Delta_{R}}$ for the three regions
of $\tau$, one can find it can be expressed in the form of
$\rho_{\blacktriangle}$ with
\begin{eqnarray}\label{SMAll}
(S,M)= \left \{
\begin{array}{lr}
\bigr((1-\gamma)/3, 0\bigr), \; &  0 \leq \gamma < \bigr( \sqrt{3}-1 \bigr) /2 ; \; \\
\bigr( \mathcal{S}(\gamma), \mathcal{M}(S) \bigr), \; &  \bigr( \sqrt{3}-1 \bigr) /2 \leq \gamma \leq 2/3; \; \\
\bigr(1-\gamma,    S\bigr), \; &  2/3 \leq \gamma  \leq 2/3. \;
\end{array}
\right.
%M= \left \{
%\begin{array}{lr}
%0,      \; &  0 \leq \gamma < \bigr( \sqrt{3}-1 \bigr) /2 \; \\
%\mathcal{M}(S),      \; &  \bigr( \sqrt{3}-1 \bigr) /2 \leq \gamma \leq 2/3 \; \\
%S,      \; &  2/3 \leq \gamma  \leq 2/3 \;,
%\end{array}
%\right.
\end{eqnarray}
This frontier state $\rho_{\Delta_{R}}$ has been verified by one
million random two-qubit states. In Fig. \ref{fig2}, we show the
curve of $\rho_{\Delta_{R}}$ in the $\Delta_{R}-\tau$ plane, in
company with 30000 random states weighted based on it.

\section{Discussion and Conclusion\label{Concl}}
In this letter, we show the properties of the fidelity lead to the
upper and lower bounds of the squared concurrence for a biparticle
system. The bounds are closely related to the schemes to observe
entanglement experimentally. The two important concepts in quantum
information, entanglement and fidelity, have been pointed out to be
related to each other in many aspects
\cite{Uhlmann2000,Wootters97,Guo,Wootters98}. Recently, the relation
between entanglement and Berry phase begins to attract the attention
of researchers \cite{ConBPhase}. On the other hand, the Uhlmann’s
mixed state geometric phase has the inherent connection with
fidelity \cite{Uhlmann2}. These suggest the bounds investigated in
this letter may induce some characters of the geometric phase of
mixed state.

As a further research of bounds of the squared concurrence, limited
to the two-qubit case, we derived the frontier states, which
corresponding to the maximal upper bound, the minimal lower bound or
the maximal difference between the bounds, for a given value of
concurrence. We also determined the state achieving the maximal
difference between the upper bound and the modified lower bound,
which was defined as the maximum of $0$ and the lower bound. All of
these frontier states were relating to this form of $\rho_{ansatz}$
in Eq. (\ref{RhoAnsatz}), a mixture of a Bell base and a mixed
diagonal state.

\begin{figure}
\includegraphics[width=7cm]{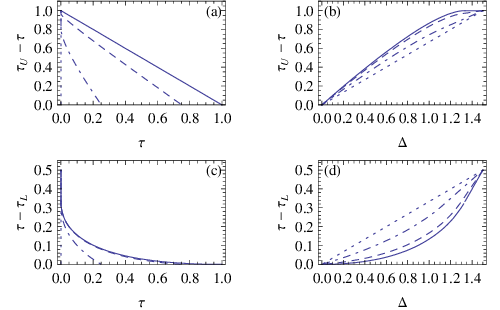} \\
 \caption{Plot of the errors of the bounds, $\tau_{U}-\tau$ and $\tau-\tau_{L}$, versus $\tau$ and $\Delta$,
 for the generalized Werner states.
 The curves show the results of $\theta=\pi/4$ (solid), $\theta=\pi/6$ (dashed), $\theta=\pi/12$ (dot-dashed) and $\theta=0$ (dotted). }
\label{fig3}
\end{figure}

We haven't discussed more general properties of the bounds in this
letter. Instead, we exhibit some characteristics, in Fig.
\ref{fig3}, by a family of generalized Werner states \cite{RhoWG},
\begin{eqnarray} \label{RhoWgeneral}
\rho_{W}(\theta)=\frac{1-x}{4}I_{2} \otimes I_{2} + x
|\Phi(\theta)\rangle \langle\Phi(\theta)|,
\end{eqnarray}
which are the mixtures of an entangled pure state
$|\Phi(\theta)\rangle = \cos \theta |00\rangle + \sin \theta
|11\rangle $ with the completely random state. One can notice, for a
given value of entanglement, both the largest errors of the upper
and lower bounds occurs when $\theta=\pi/4$, corresponding with the
originally defined Werner state in Eq. (\ref{RhoW}). When the linear
entropy (equalling to $\Delta$ here) increases, the errors are
enhanced monotonously. These results provided the preparatory
knowledge to derive the frontier states in Sec. \ref{Frontier}.

\begin{acknowledgments}
This work is supported by the New teacher Foundation of Ministry of
Education of P.R.China (Grant No. 20070248087). J.L.C is supported
in part by NSF of China (Grant No. 10605013), and Program for New
Century Excellent Talents in University, and the Project-sponsored
by SRF for ROCS, SEM.
\end{acknowledgments}

%\begin{thebibliography}{99}
\bibliography{ConFid_flzhang_rsb}

%\end{thebibliography}
\end{document}